\documentclass[9pt, twocolumn]{article}
\usepackage{times,bm}
\usepackage[affil-it,]{authblk}
\usepackage[pagebackref=false,colorlinks=true,linkcolor=acsblue,citecolor=red,urlcolor=acsblue]{hyperref}
\usepackage{geometry}
\usepackage[switch*, displaymath,pagewise, mathlines]{lineno}
\usepackage{graphicx}
\usepackage{cite}
\usepackage{amsmath}
\usepackage{amsfonts}
\usepackage{amssymb}
\usepackage{slashed}
\usepackage{subfloat}
\usepackage{slashed}
\usepackage{color}
\usepackage{float}
\usepackage{multirow,multicol}
\usepackage{tabulary}
\usepackage[flushleft]{threeparttable}
\usepackage{pgfplots,subfigure}
\usepackage{xcolor}
\numberwithin{equation}{section}
\usepackage{tikz}
\usepackage{ulem}
\usepackage{hyperref}
\usepackage{nameref}
\usepackage{comment}
\usepackage{xcolor}
\definecolor{acsblue}{RGB}{17,76,139}
\definecolor{shadecolor}{RGB}{255,241,204}

\usepgflibrary{arrows}
\usetikzlibrary{shapes.callouts}
\tikzset{
	level/.style   = { thick, },
	connect/.style = { dotted, red   },
	notice/.style  = { draw, rectangle callout, callout relative pointer={#1} },
	label/.style   = { text width=2cm }
}
\usepackage{letltxmacro}
\makeatletter
\let\oldr@@t\r@@t
\def\r@@t#1#2{%
	\setbox0=\hbox{$\oldr@@t#1{#2\,}$}\dimen0=\ht0
	\advance\dimen0-0.2\ht0
	\setbox2=\hbox{\vrule height\ht0 depth -\dimen0}%
	{\box0\lower0.4pt\box2}}
\LetLtxMacro{\oldsqrt}{\sqrt}
\renewcommand*{\sqrt}[2][\ ]{\oldsqrt[#1]{#2}}
\makeatother
\usepackage{environ}
\begin{document}
\def\nofundrefquery{}
\def\stmdocstextcolor#1{}
\def\stmdocscolor#1{}
\newcommand{{\ri}}{{\rm{i}}}
\newcommand{{\Psibar}}{{\bar{\Psi}}}
\renewcommand{\rmdefault}{ptm}

\title{\mdseries{Photonic Modes in Twisted Graphene Nanoribbons}}

\author{ \textit {\mdseries{Abdullah Guvendi}}$^{\ 1}$\footnote{\textit{ E-mail: abdullah.guvendi@erzurum.edu.tr  } }~,~ \textit {\mdseries{Semra Gurtas Dogan}}$^{\ 2}$\footnote{\textit{ E-mail: semragurtasdogan@hakkari.edu.tr (Corr. Author) } }~,~ \textit {\mdseries{Omar Mustafa}}$^{\ 3}$\footnote{\textit{ E-mail: omar.mustafa@emu.edu.tr} }~,~ \textit {\mdseries{Kobra Hasanirokh}}$^{\ 4}$\footnote{\textit{ E-mail: zhasanirokh@yahoo.com } }  \\
	\small \textit {$^{\ 1}$Department of Basic Sciences, Erzurum Technical University, 25050, Erzurum, Turkiye}\\
	\small \textit {$^{\ 2}$Department of Medical Imaging Techniques, Hakkari University, 30000, Hakkari, Turkiye}\\
	\small \textit {$^{\ 3}$Department of Physics, Eastern Mediterranean University, 99628, G. Magusa, north Cyprus, Mersin 10 - Turkiye}\\
 \small \textit {$^{\ 4}$Research Institute for Applied Physics and Astronomy, University of Tabriz, Tabriz 51665-163, Iran}}
\date{}
\maketitle

\begin{abstract}
This study investigates the behavior of photonic modes in twisted graphene nanoribbons (TGNRs) using an analytical approach based on solving the fully covariant vector boson equation. We present a model that demonstrates how helical twisting in TGNRs significantly affects the evolution of photonic modes. Our analytical solutions yield detailed expressions for mode profiles, energy spectra, and decay characteristics. We find that increasing the twist parameter shortens the decay times (\(\tau_{ns}\)) for damped modes, indicating enhanced photonic coupling due to the twisted geometry. Conversely, longer nanoribbons (NRs) exhibit increased decay times, showing a length (\(L\))-dependent effect, where \(\tau_{ns} \propto L / c\), with \(c\) representing the speed of light. These findings may enhance the understanding of light control in nanostructures and suggest potential applications in tunable photonic devices, topological photonics, and quantum optical systems.
\end{abstract}

\begin{small}
\begin{center}
\end{center}
\end{small}
\section{\mdseries{Introduction}}\label{sec1}

Graphene \cite{graphene}, a two-dimensional material renowned for its extraordinary mechanical strength, electrical conductivity, and optical transparency, has become a foundational element in the study of nanoscale physics and materials science \cite{k1,k2}. Graphene nanoribbons (GNRs), narrow strips of graphene, introduce quantum confinement and edge effects that further enhance graphene’s already exceptional properties, making them particularly attractive for applications in nanoscale electronics, optoelectronics, and photonics \cite{k3,bonaccorso2010graphene}. Among the various configurations of GNRs, TGNRs have recently garnered considerable attention due to their ability to manipulate photonic and electronic properties through geometric deformation \cite{a1,a2}. The twist introduces a helical structure into the NR, which can significantly alter the behavior of electromagnetic modes confined within the material. This deformation provides new degrees of freedom for controlling light-matter interactions, enabling novel functionalities in nanophotonic devices \cite{a3}. The helical nature of TGNRs modifies the boundary conditions \cite{a4} imposed on photonic modes \cite{a5}, leading to shifts in the dispersion relations, mode confinement, and decay dynamics. In particular, TGNRs have been shown to support chiral edge modes, which are of significant interest for topological photonics \cite{a6}, where light is protected from backscattering by defects and impurities. These topologically protected states are not only robust but also allow for unidirectional light transport, making them promising candidates for applications in quantum information processing and photonic circuitry \cite{a7}. The introduction of a helical geometry also leads to the emergence of geometric potentials that influence the photonic band structure \cite{a8}. These potentials can open gaps in the spectrum \cite{a9}, which are associated with topologically protected edge states. Such edge states offer exciting opportunities for the development of photonic devices that are robust against material imperfections and disorder. Furthermore, the ability to tune the twist parameter provides a mechanism to control the photonic mode lifetimes, allowing for the design of devices with adjustable decay rates. However, a theoretical framework that fully captures the covariant nature of electromagnetic fields in these twisted nanostructures has been lacking.

On the other hand, Barut introduced a comprehensive framework for systematically deriving the well-known fully-covariant wave equations that describe the dynamics of spinning particles \cite{barut}. In this framework, the vector boson equation emerges as an excited state of zitterbewegung, specifically related to the spin-1 sector of the Duffin-Kemmer-Petiau equation in (2+1) dimensions. The corresponding spinor is formulated as a direct product of two symmetric Dirac spinors, resulting in a symmetric rank-two spinor within the vector boson equation \cite{barut}. This method allows for the extraction of non-perturbative results applicable to a range of physical systems. It has also been demonstrated that the Duffin-Kemmer-Petiau equation is equivalent to the complex Proca equation, and in the massless limit, the Proca equation simplifies to the Maxwell equation \cite{nuri1,nuri2}.  Various research avenues within this context are noteworthy, such as the quantum analog of Schumann resonances, the study of quantum tunneling for spin-1 particles from Warped-AdS$_{3}$ black holes \cite{ganim}, the evolution of relativistic spin-1 oscillator fields near black hole horizons \cite{as1}, the behavior of vector bosons in non-zero cosmological settings \cite{as2}, and evolution of photons within the rotating frame of negative curvature wormholes \cite{as3}. Nonetheless, there has been no exploration of photons in helicoidal backgrounds or twisted nanomaterials. To address this gap, we analytically solve the fully covariant vector boson equation to investigate photonic modes in TGNRs. This approach leads to a Schrödinger oscillator-like wave equation, which facilitates the determination of energy profiles, mode evolution, and decay properties of photonic states in these twisted structures. 

This manuscript is structured as follows: In Section \ref{sec2}, we revisit the spacetime background that describes twisted helicoidal structures. Section \ref{sec3} presents a non-perturbative wave equation that captures the dynamics of photons in TGNRs and provides analytical solutions for their behavior within these structures. Our analysis reveals that the photonic modes in TGNRs are intrinsically unstable and experience decay over time due to energy loss. We show that the decay time of these modes is affected by both the total length of the NR and the number of twists. Specifically, our model predicts that the decay time of photonic modes could range from \(10^{-19}\) to \(10^{-16}\) seconds, depending on whether the NR length varies from 1 nm to \(10^3\) nm. Finally, Section \ref{sec4} provides a summary and discusses the implications of our results.

\section{\mdseries{Revisiting the TGNRs background}} \label{sec2}

The helicoidal background describing the TGNRs can be characterized through the parametrization \cite{a8} 
\begin{eqnarray}
\vec{r}\left(u,\nu\right)=\nu\hat{i}+u \,cos(\omega v)\hat{j}+u\, sin(\omega v)\hat{k},
\end{eqnarray}
where \( v\in[-\frac{L}{2},\frac{L}{2}] \) and \( u\in[-\frac{D}{2},\frac{D}{2}] \). Here \( D \) is the width of the NR and \( L \) is the total length of the NR, which is aligned around the \( x \)-axis. The parameter \( w=\frac{2\pi }{L}m \), a real number with units of inverse length, determines the chirality of the surface (twist parameter), and \( m \) is the number of \( 2\pi \) twists. Given this parametrization, one finds
\begin{equation*}
\begin{split}
&dx = dv,\\
&dy = \frac{\partial y}{\partial u} du + \frac{\partial y}{\partial v} dv \Rightarrow \cos(w \nu) du - u w \sin(w v) dv,\\
&dz = \frac{\partial z}{\partial u} du + \frac{\partial z}{\partial v} dv \Rightarrow \sin(w v) du + u w \cos(w \nu) dv,\\
\end{split}
\end{equation*}
Accordingly, we have
\begin{equation*}
dx^2 + dy^2 + dz^2 \Rightarrow du^2 + \left(1 + u^2 w^2\right) dv^2.
\end{equation*}
By trivially projecting the temporal coordinate (\( t \)) from flat space-time, where the helicoidal ribbon resides, we can describe the TGNRs (i.e., helicoidal surface) through the following \((2+1)\)-dimensional curved spacetime metric \cite{a8} 
\begin{eqnarray}
ds^2=c^2dt^2-du^2-\chi\left(u\right)dv^2, \label{eq1}
\end{eqnarray}
where \( \chi\left(u\right)=1+w^2u^2 \), and \( c \) is the speed of light in vacuum. Based on the line element in Eq. (\ref{eq1}), the covariant metric tensor can be expressed as
\begin{equation*}
g_{\mu \nu}=\text{diag}\left(c^2, -1, -\chi\left(u\right)\right).
\end{equation*}
The space-independent (free) Dirac matrices are represented using the Pauli spin matrices (\(\sigma^{x}, \sigma^{y}, \sigma^{z}\)) in the following manner: \( \gamma^{0}=\sigma^{z} \), \( \gamma^{1}=i\sigma^{x} \), \( \gamma^{2}=i\sigma^{y} \), which aligns with the signature (\(+,-,-\)) of the line element \cite{as4,g-btz}. The spinorial affine connections for the Dirac field, denoted \( \Gamma_{\lambda} \), are given by: \(\Gamma_{\lambda} = \frac{1}{4} g_{\mu \tau} \left[e^{k}_{\nu,\lambda} e^{\tau}_{k} - \Gamma_{\nu \lambda}^{\tau} \right] \mathcal{S}^{\mu \nu}  \)\cite{as3}, where $,\lambda$ denotes differentiation with respect to \( x^{\lambda} \). Here, \( \Gamma_{\nu \lambda}^{\tau} \) represents the Christoffel symbols, defined as: \( \Gamma_{\nu \lambda}^{\tau} = \frac{1}{2} g^{\tau \epsilon} \left[\partial_{\nu} g_{\lambda \epsilon} + \partial_{\lambda} g_{\epsilon \nu} - \partial_{\epsilon} g_{\nu \lambda} \right] \)\cite{as3}. The inverse tetrad fields are denoted by \( e^{\tau}_{k} \), and \( \mathcal{S}^{\mu \nu} \) represents the spin operators, defined by: \( \mathcal{S}^{\mu \nu} = \frac{1}{2} \left[\gamma^{\mu}, \gamma^{\nu} \right] \) \cite{as3,as4}. Greek indices refer to the coordinates in the curved spacetime, while Latin indices refer to the coordinates in flat Minkowski spacetime. The tetrad fields (and their inverses \( e^{\mu}_{k} \)) are derived as follows:
\begin{flalign*}
e^{k}_{\mu}=\left(\begin{array}{ccc}
c & 0 & 0\\
0 & 1 & 0\\
0 & 0 & \sqrt{\chi(u)}
\end{array}\right), e^{\mu}_{k}=\left(\begin{array}{ccc}
\frac{1}{c} & 0 & 0\\
0 & 1 & 0\\
0 & 0 & \frac{1}{\sqrt{\chi(u)}}
\end{array}\right)
\end{flalign*}
since \( g_{\mu\nu}=e^{k}_{\mu}e^{l}_{\nu}\eta_{kl} \) and \( e^{\mu}_{k}=g^{\mu\nu}e^{l}_{\nu}\eta_{kl} \) where \( \eta_{kl} \) is the Minkowski tensor, \( \eta_{kl}=\textrm{diag}\left(1,-1,-1\right) \) \cite{as4}. Accordingly, we have \( \gamma^{t}= \frac{1}{c}\sigma^{z} \), \( \gamma^{u}= i \sigma^{x} \), \( \gamma^{v}=\frac{1}{\sqrt{\chi(u)}}i \sigma^{y} \), where \( i=\sqrt{-1} \). Also, non-zero components of the Christoffel symbols are found as follows: \( \Gamma_{vv}^{u}=-u w^2 \), \( \Gamma_{v u}^{v}=\Gamma_{u v}^{v}=\frac{u w^2}{\chi(u)} \). Additionally, the non-zero component of the spinorial affine connection is obtained as \( \Gamma_{v}=\frac{u w^2}{2\sqrt{\chi(u)}} i \sigma^{z} \).

\section{\mdseries{Wave equation and its solutions}}\label{sec3}

\begin{figure}[ht]
\centering
\includegraphics[scale=0.60]{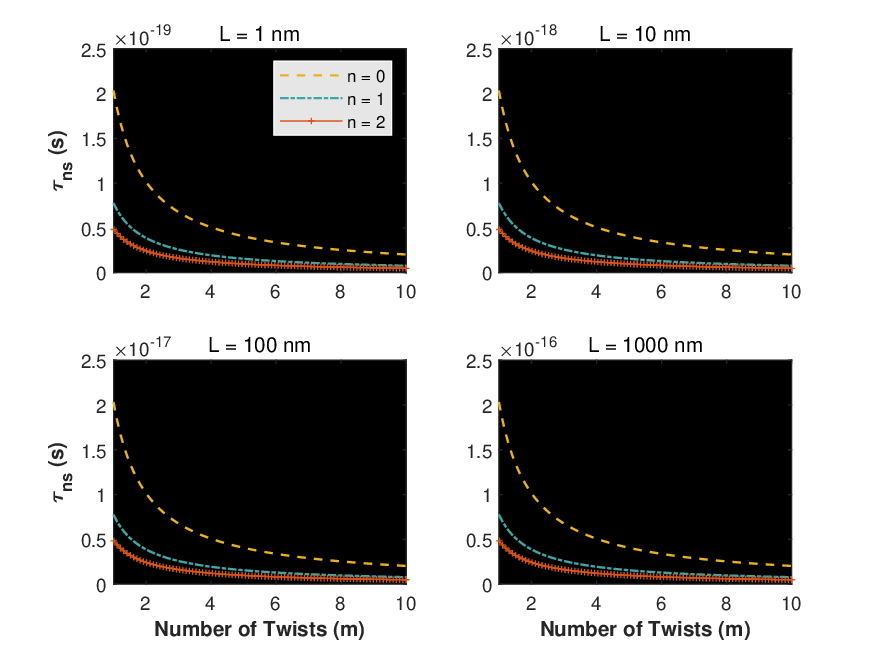}
\caption{Decay time ($\tau_{ns}$) of photonic modes as a function of the number of twists ($m$) in the NR. The total length of the NR is denoted by $L$, and we set $s=1$.}
\label{fig:1}
\end{figure}

In this section, we introduce the generalized vector boson equation for a $(2+1)$-dimensional curved spacetime. We then derive coupled equations for relativistic spin-1 vector bosons within TGNRs and seek analytical solutions for the resulting wave equation. The fully-covariant vector boson equation in $(2+1)$-dimensional curved spacetime is expressed as \cite{as3}:
\begin{equation}
\left(\mathcal{B}^{\mu} \slashed{\nabla}_{\mu} + i \tilde{m} \textbf{I}_4\right)\Psi(x^{\mu})=0,\label{VBE}
\end{equation}
where Greek indices $(\mu=t,u,v)$ refer to the curved spacetime coordinates, and $\slashed{\nabla}_{\mu}$ denotes the covariant derivative, with $\slashed{\nabla}_{\mu} = \partial_{\mu} - \Omega_{\mu}$. The matrices $\mathcal{B}^{\mu}$ are space-dependent spin-1 matrices derived from the generalized Dirac matrices $(\gamma^{\mu})$ and are expressed as \cite{as3}:
\begin{equation*}
\mathcal{B}_{\mu} = \frac{1}{2} (\gamma^{\mu} \otimes \textbf{I}_{2} + \textbf{I}_{2} \otimes \gamma^{\mu}),
\end{equation*}
where $\tilde{m} = \frac{mc}{\hbar}$, with $m$ representing the rest mass of the vector boson, and $\hbar$ the reduced Planck constant. The vector $x^{\mu}$ denotes the spacetime position vector. Here, $\textbf{I}_2$ and $\textbf{I}_4$ are the $2$-dimensional and $4$-dimensional identity matrices, respectively. The spinorial affine connections for the spin-1 field, $\Omega_\mu$, are determined from the affine spin connections $(\Gamma_\mu)$ for the Dirac fields as \cite{as3}:
\begin{equation*}
\Omega_{\mu} = \Gamma_{\mu} \otimes \textbf{I}_{2} + \textbf{I}_{2} \otimes \Gamma_{\mu}.
\end{equation*}
Additionally, $\Psi$ represents the symmetric rank-two spinor constructed as the direct product of two symmetric Dirac spinors and can be factorized as follows \cite{as3}:
\begin{equation*}
\Psi(x^{\mu})=\exp(-i\tilde{\mathcal{E}}\,t)\exp(isv)\left(\psi_{1}(u),\psi_{2}(u),\psi_{3}(u),\psi_{4}(u)\right)^{T},
\end{equation*}
where $\tilde{\mathcal{E}}=\frac{E}{\hbar c}$, $E$ is the relativistic energy, $s$ is the spin, and $^{T}$ denotes the transpose of the $u$-dependent spinor. Accordingly, one derives the following set of equations after some algebra (see also \cite{as3}):
\begin{equation}
\begin{split}
&\tilde{\mathcal{E}}\phi_{1}(u)-\tilde{m}\phi_{2}(u)-\frac{s}{\sqrt{\chi(u)}}\phi(u)=0,\\
&\tilde{\mathcal{E}}\phi_{2}(u)-\tilde{m}\phi_{1}(u)-\dot{\phi}(u)=0,\\
&\tilde{m}\phi(u)+\dot{\phi_{1}}(u)+\frac{uw^2}{\chi(u)}\phi_{1}(u)-\frac{s}{\sqrt{\chi(u)}}\phi_{2}(u)=0,\label{Eq-set}
\end{split}
\end{equation}
where the dot indicates the derivative with respect to the variable, $\phi_{1}(u)=\psi_{1}(u)+\psi_{4}(u)$, $\phi_{2}(u)=\psi_{1}(u)-\psi_{4}(u)$, and $\phi(u)=\psi_{2}(u)+\psi_{3}(u)$ since $\psi_{2}(u)=\psi_{3}(u)$. Solving this set of equations for $\phi(u)$, we obtain the following wave equation
\begin{equation}
\ddot{\phi}(u)+\frac{uw^2}{\chi(u)}\dot{\phi}(u)+\left[\tilde{\mathcal{E}}^2-\frac{s^2}{\chi(u)}\right]\phi(u)=0,\label{om1}
\end{equation}
for photons. We now use the change of variables in the form of  \( u=\frac{sinh(x)}{w} \in \mathbb{R}\) to imply that (\ref{om1}) transforms into 
\begin{equation}
\ddot{\phi}(x)+\left[\frac{\tilde{\mathcal{E}}^2}{w^2}\,\cosh^{2}(x)-s^2\right]\phi(x)=0.\label{om2}
\end{equation}
This wave equation precisely characterizes the photonic modes in twisted graphene nanostructures. However, finding an exact analytical solution to this equation does not seem feasible. At this point, assuming that the NR is narrow, solutions can be obtained with a good approximation. Under the approximation \(\cosh^2(x) \approx 1+{x}^{2}+\mathcal{O} \left( {x}^{4} \right)\), the wave equation simplifies to the well known one-dimensional Schrödinger inverted-oscillator form \cite{OM1,OM2}
\begin{equation}
\ddot{\phi}(x)+\left[kx^2+\tilde{a}\right]\phi(x)=0, \label{om3}
\end{equation}
where $\tilde{a}=k-s^2,\quad k=\tilde{\mathcal{E}}^2/w^2.$ We may now follow the textbook procedure and use the variable $y=\sqrt{k}x^2$ to obtain
\begin{equation}
y\,\ddot{\phi}(y)+\frac{1}{2}\dot{\phi}(y)+\left[\frac{y}{4k}+\frac{\tilde{a}}{4k}\right]\phi(y)=0 . \label{om4}
\end{equation}
At this point, we may use power series expansion (following a standard textbook procedure starting with the substitution $  \phi (y)=e^{-iy/2}\, U(y),$ to yield 
\begin{equation}
y\,\ddot{U}(y)-\left[iy -\frac{1}{2}\right]\dot{U}(y)-\left[\frac{i}{4}-\frac{\tilde{a}}{4\sqrt{k}}\right]U(y)=0 . \label{om5}
\end{equation}
We may now use$$ U(y)=\sum_{j=0}^{\infty }A_{j}\,y^{j+\sigma },$$ to obtain%
\begin{equation}
\begin{split}
 & \sum_{j=0}^{\infty }A_{j}\,\left[ \frac{\tilde{a}}{4\sqrt{k}}-\frac{i}{4}-i(j+\sigma)\right]\,y^{j+\sigma } \\
 &+  \sum_{j=0}^{\infty }A_{j}\, \left[(j+\sigma)(j+\sigma-\frac{1}{2})\right]\,y^{j+\sigma -1}=0. 
  \end{split}
\end{equation}
This would, in turn, imply the two terms recursion relation
\begin{equation}
\begin{split}
  \sum_{j=0}^{\infty }&\left(A_{j}\,\left[ \frac{\tilde{a}}{4\sqrt{k}}-\frac{i}{4}-i(j+\sigma)\right] \right.\\
  &\left.+A_{j+1}\, \left[(j+\sigma+1)(j+\sigma+\frac{1}{2})\right]\right)\,y^{j+\sigma }\\
 &+A_0\,\sigma(\sigma-\frac{1}{2})\,y^{\sigma-1} =0.
  \end{split} \label{om6}
\end{equation}
$A_0\neq0\Rightarrow\,\sigma=0,1/2$. We take $\sigma=1/2$ so that $U(y)\rightarrow 0$ as $y\rightarrow 0$. Consequently, 
\begin{equation}
\begin{split}
& A_{j}\,\left[ \frac{\tilde{a}}{4\sqrt{k}}-\frac{i}{4}-i(j+\frac{1}{2})\right]+A_{j+1}\, \left[(j+\frac{3}{2})(j+1)\right] =0. 
\end{split}\label{om7}
\end{equation}
The power series is truncated to a polynomial of order $n\geq0$ by the requirement that $\forall j=n$,  $A_{n+1}=0$ and $A_{n}\neq0$. This would manifestly suggest that 
\begin{equation}
    \frac{\tilde{a}}{4\sqrt{k}}-\frac{i}{4}-i(n+\frac{1}{2})=0 \Rightarrow   \tilde{a}{}=4i\sqrt{k}\tilde{n}, \label{om8}
\end{equation}
where $\tilde{n}=n+3/4$. Hereby, one should observe that this result is in exact accord with that of the two-dimensional inverted Schr\"{o}dinger oscillator  (taking into account the units $\hbar=2m=1$ in the Schr\"{o}dinger equation in \cite{OM1} and the isomorphism between the angular momentum quantum number $\ell_d$ and dimensionality $d$ discussed and detailed  in, e.g., \cite{OM3}).  Hence,
\begin{equation}
\begin{split}
&k-4i\sqrt{k}\tilde{n}-s^2=0\Rightarrow [\sqrt{k}-2i\tilde{n}]^2=(2i\tilde{n})^2+s^2 \\
&\Rightarrow \tilde{\mathcal{E}}=-iw\left[2\tilde{n}+\sqrt{4\tilde{n}^2-s^2}\right]\\
&\Rightarrow E_{ns}=-i\hbar c w\left[2\tilde{n}+\sqrt{4\tilde{n}^2-s^2}\right].
\end{split}\label{om9}
\end{equation}
Moreover, our $\phi_{ns}(y)$ now reads
\begin{equation*}
\phi_{ns}(y)=\sqrt{y}\, e^{iy/2}\sum_{j=0}^{n}A_j y^j,
\end{equation*}
where $ A_0=1$ and
\begin{equation*}
A_{j+1}=\frac{\left[ \frac{\tilde{a}}{4\sqrt{k}}-\frac{i}{4}-i(j+\frac{1}{2})\right]}{\left[(j+\frac{3}{2})(j+1)\right]}A_{j}\,; \,0\leq j\leq n,
\end{equation*}
to identify the power series coefficients. It could be interesting to mention that the above procedure has been successfully used in, e.g., \cite{OM4,OM5,OM6,OM7}. One should, nevertheless, observe that our result in (\ref{om9}) manifestly introduces a decay time (since $\Psi\propto \exp(-i\tilde{\mathcal{E}}_{ns}t)$), of  the photonic modes is given by:
\begin{equation}
\tau_{ns}=\frac{\hbar}{|\Im E_{ns}|}\Rightarrow \tau_{ns}=\frac{(L/c)}{2\pi m\,\left[2\tilde{n}+\sqrt{4\tilde{n}^2-s^2}\right]}.\label{ES}
\end{equation}
This result clearly indicates that while the decay time of a given $ns$-photonic mode is directly proportional to the  total length of the NR ($L$), it is inversely proportional to the number of $2\pi$ twists ($m$). One should also observe that larger $n$  photonic modes decay more rapidly than the photonic modes with smaller $n$ values. These effects are made clear in Figure \ref{fig:1}. Yet, if we consider the usual value of the speed of light, our model predicts that the decay time of photonic modes could range from \(10^{-19}\) to \(10^{-16}\) seconds if the $L$ varies from 1 nm to \(10^3\) nm. At first look, we can see that the order of the decay time is determined by the ratio $L/c$.
\section{\mdseries{Summary and discussions}}\label{sec4}

In this manuscript, we present a model for the evolution of photonic modes in TGNRs. We start by deriving a spacetime background that characterizes the helicoidal geometry of the twisted surface and solve the corresponding fully covariant vector boson equation analytically. This approach yields a non-perturbative wave equation for photons within the TGNRs. To simplify the analysis, we use the approximation \(\cosh^2(x) \approx 1 + x^2\), where \(x\) relates with the material's width. We find that the photonic modes are inherently unstable and decay over time. The decay time is affected by several factors: the NR's total length, the number of twists, the speed of light in the material, and the quantum numbers \(s\) (spin) and \(n\) (energy level). Using the standard speed of light in a vacuum, we estimate that the decay times of these modes range from \(10^{-19}\) to \(10^{-16}\) seconds for NRs with total lengths between 1 nm and 1000 nm. Our study advances the understanding of how geometry and topology affect light behavior in low-dimensional systems, particularly in TGNRs. By solving the fully covariant vector boson equation, we gain new insights into the evolution of photonic modes in these materials. Our findings suggest that both the confinement and decay characteristics of photonic modes can be controlled by adjusting the twist and total length of the NRs.

Damped photonic modes within TGNRs can be closely linked to resonant states, where photons become confined or exhibit oscillatory behavior while progressively losing energy over time. These resonances typically arise at specific frequencies or wavelengths where the nanomaterial exhibits enhanced interactions with the electromagnetic field, leading to pronounced resonance effects. The damping observed is a direct consequence of energy dissipation mechanisms such as photon absorption. These findings offer significant potential for advancing photonic technologies. By leveraging the geometric versatility of nanomaterials, it is possible to engineer optical properties with high precision. This capability paves the way for innovative applications, such as tunable laser systems, highly efficient waveguides, and quantum optical devices, where precise control of light-matter interactions is essential. Specifically, the ability to manipulate photonic mode confinement and tailor decay rates through structural modifications in TGNRs could enable the development of customizable and highly responsive optical systems.

Our results underscore the importance of geometric design in influencing the photonic behavior of low-dimensional materials. By optimizing structural parameters, it becomes feasible to control resonance phenomena, energy dissipation, and mode confinement. Thus, this work can be regarded as a significant step toward the integration of advanced nanophotonic architectures in both applied and fundamental research domains, with profound implications for future technological breakthroughs in areas such as quantum computing, optical communications, and beyond.
\section*{\small{Data availability}}
This is a theoretical research, including equations and derived quantities presented in the main text of the paper.
\section*{\small{Conflicts of interest statement}}
The authors have disclosed no conflicts of interest.
\section*{\small{Funding}}
This research has not received any funding.

\end{document}